\documentclass[a4paper]{article}
\usepackage{hyperref}
\hypersetup{breaklinks=true,colorlinks,citecolor=blue,linkcolor=blue}
\usepackage{INTERSPEECH2020}
\usepackage{graphicx,amsmath,amsthm,amssymb}
\usepackage{url,color}
\usepackage{algorithm,algorithmic}
\graphicspath{{grf/}}
\title{InfantNet: A Deep Neural Network for Analyzing Infant Vocalizations}
\name{Mohammad K.~Ebrahimpour$^1$, Sara Schneider$^2$, David C.~Noelle$^{1,2}$ and Christopher T.~Kello$^2$}

\address{
  $^1$Electrical Engineering and Computer Science, UC Merced\\
  $^2$Cognitive and Information Sciences, UC Merced}
\email{\{mebrahimpour,sschneider2,dnoelle,ckello\} @ucmerced.edu}

\begin{document}

\maketitle
\begin{abstract}
Acoustic analyses of infant vocalizations are valuable for research on speech development as well as applications in sound classification. Previous studies have focused on measures of acoustic features based on theories of speech processing, such spectral and cepstrum-based analyses. More recently, end-to-end models of deep learning have been developed to take raw speech signals (acoustic waveforms) as inputs and convolutional neural network layers to learn representations of speech sounds based on classification tasks. We applied a recent end-to-end model of sound classification to analyze a large-scale database of labeled infant and adult vocalizations recorded in natural settings outside the lab with no control over recording conditions. The model learned basic classifications like infant versus adult vocalizations, infant speech-related versus non-speech vocalizations, and canonical versus non-canonical babbling. The model was trained on recordings of infants ranging from 3 to 18 months of age, and classification accuracy changed with age as speech became more distinct and babbling became more speech-like. Further work is needed to validate and explore the model and dataset, but our results show how deep learning can be used to measure and investigate speech acquisition and development, with potential applications in speech pathology and infant monitoring. 
\end{abstract}

\section{Introduction}
Infants produce cries and other non-speech vocalizations from birth, and they start to produce more complex, speech-like vocalizations as early as 3 months of age~\cite{infant1,infant2}. They enter a period of exploring speech-like vocalizations known as babbling, and they may start speaking their first words around one year of age. But even before then, vocalizations carry information about physical and emotional states that matter to caregivers and infant health and well-being~\cite{infant1,infant2}. Changes in vocalizations over time carry information about speech development that may be valuable for informing theories and diagnosing when development is atypical. 

Studies of infant vocalizations often borrow analysis techniques formulated for speech analysis, such as mel-frequency cepstral coefficents~\cite{survey,logmel,log-mel}. These techniques rely on an engineered feature space based in part on theories of human speech perception and production. One of the most powerful types of engineered representation for speech recognition tasks is based on the mel-frequency cepstrum~\cite{log-mel}, which is basically the discrete cosine transform of the windowed spectra. Researchers have used such engineered features as inputs to machine learning models for automatic speech recognition (ASR)~\cite{ASR} and music classification~\cite{music}. In these cases, inputs are typically two-dimensional feature maps created by arranging the log-mel cepstral features of each frame along the time axis. This feature map creates locality in both time and frequency domains~\cite{10}, which means that the machine learning problem can be framed as an image classification problem.

With respect to classifying infant vocalizations, Rosita and Junaedi developed a system using MFCC features based on voice type~\cite{rosita}. They classified vocalizations into categories of hungry, discomfort, and tiredness ~\cite{rosita}, and found MFCC features are well suited for discriminating these categories. In another study, Alaie and colleagues used Gausian Mixture Models to classify healthy and sick infants based on infant cries using MFCC features with promising results~\cite{alaie}. Similarly, Zabidi et~al. proposed a multi-layer perceptron to classify infant cries with Asphyxia~\cite{zabidi}. 

While models have progressed to date based on traditional speech analysis techniques, it is worth noting that pre-linguistic vocalizations are often atypical relative to adult speech. More generally, acoustic-based features that are tailored to infant vocalizations and distinctions may be better suited for classification tasks and research on speech development. In recent years, advances in deep neural networks have given rise to so-called ''end-to-end" models that take mostly unprocessed audio and video signals as inputs, and learn layers of features that represent the signals along dimensions that are helpful for any given classification task. 

In particular, convolutional neural networks (CNNs) have proven effective in learning to classify large sets of categories when given very large numbers of training examples~\cite{vgg,resnet,VDNet,ebrahimpour_sensitivity}. One of the advantages of deep CNNs in sound classification is their ability to learn useful features in an end-to-end manner by mapping raw data, such as raw waveform audio, onto class labels.
For instance, Dai et~al. proposed 5 CNNs with different architectures and a varying number of parameters~\cite{day_very}. AclNet~\cite{aclnet} is another end-to-end CNN architecture, inspired by MobileNet~\cite{mobilenet} because of its computational efficiency. AclNet achieved human-level accuracy for the ESC50 dataset with only 155k parameters and 49.3 million multiply-adds per second~\cite{aclnet}. 

In the present study, we take a data-driven approach classifying infant vocalizations by applying an end-to-end deep learning model of sound classification to a database of infant and adult vocalizations~\cite{ebrahimpour_icassp}. Vocalizations recorded in natural settings served as inputs in the form of raw waveforms to convolutional network layers that learned representations useful for classifying vocalizations. We also tested whether an ''Inception Nucleus" layer might improve performance by providing more varied convolutional filters. This and other features of our model architecture were designed to minimize the number of parameters and thereby avoid over-fitting to training data.

\section{Infant Vocalization Dataset}
In the present study, we trained the sound classification model introduced by Ebrahimpour et~al.~\cite{ebrahimpour_icassp} on a database of sound recordings collected by Warlaumont and colleagues~\cite{pretzer2019infant}. They recruited families with infants to participate in a study of infant speech development. Each family agreed to put a vest on their infant that had a pocket containing an audio recording device (LENA;~\cite{lena}). The infant wore the vest for 12-hour periods at 3, 6, 9, and 18 months of age. Parents were advised to record during the weekend when they were most likely to be together with the infant, and the parents were compensated for their participation (\$20 at 3-months, \$30 at 6-months, \$40 at 9-months, \$60 at 18-months, and an additional \$40 and a summary of the output from their recordings at the end of the study). On recording days, families otherwise went about their business as usual, and recordings captured all sounds near the infant, including their vocalizations, nearby adult vocalizations, and various other ambient sounds. 

Three 5-minute periods of relatively active vocalizations by the infant were identified for each day-long recording with assistance of the LENA system. Research assistants transcribed each 5-minute period to identify time segments that contained infant and adult vocalizations. Infant vocalizations were categorized as laugh/cry (sometimes both in the same vocalization period), canonical babbling (well-structured syllables with consonants and vowels), and non-canonical babbling (lone vowels or consonants, less speech-like sounds). Adult vocalizations were categorized as either directed at the infant, such as vocal play meant to elicit responses, or directed at other adults or other children nearby. Any sensitive information (e.g. credit card numbers) that happened to be recorded was removed.

\begin{figure}
    \centering
    \begin{tabular}{cc}
        \includegraphics[width= 0.5 \linewidth]{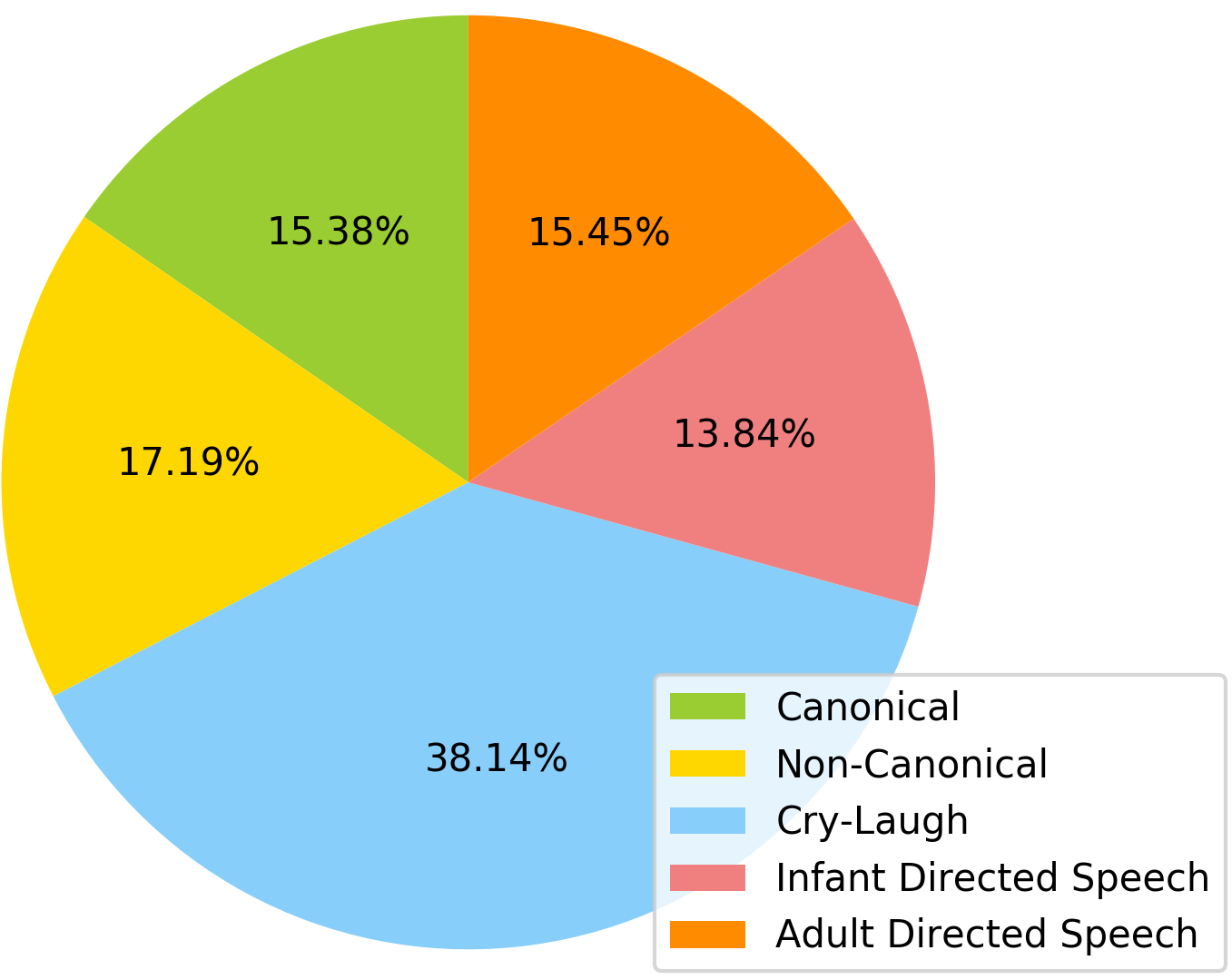}&
        \includegraphics[width= 0.42 \linewidth]{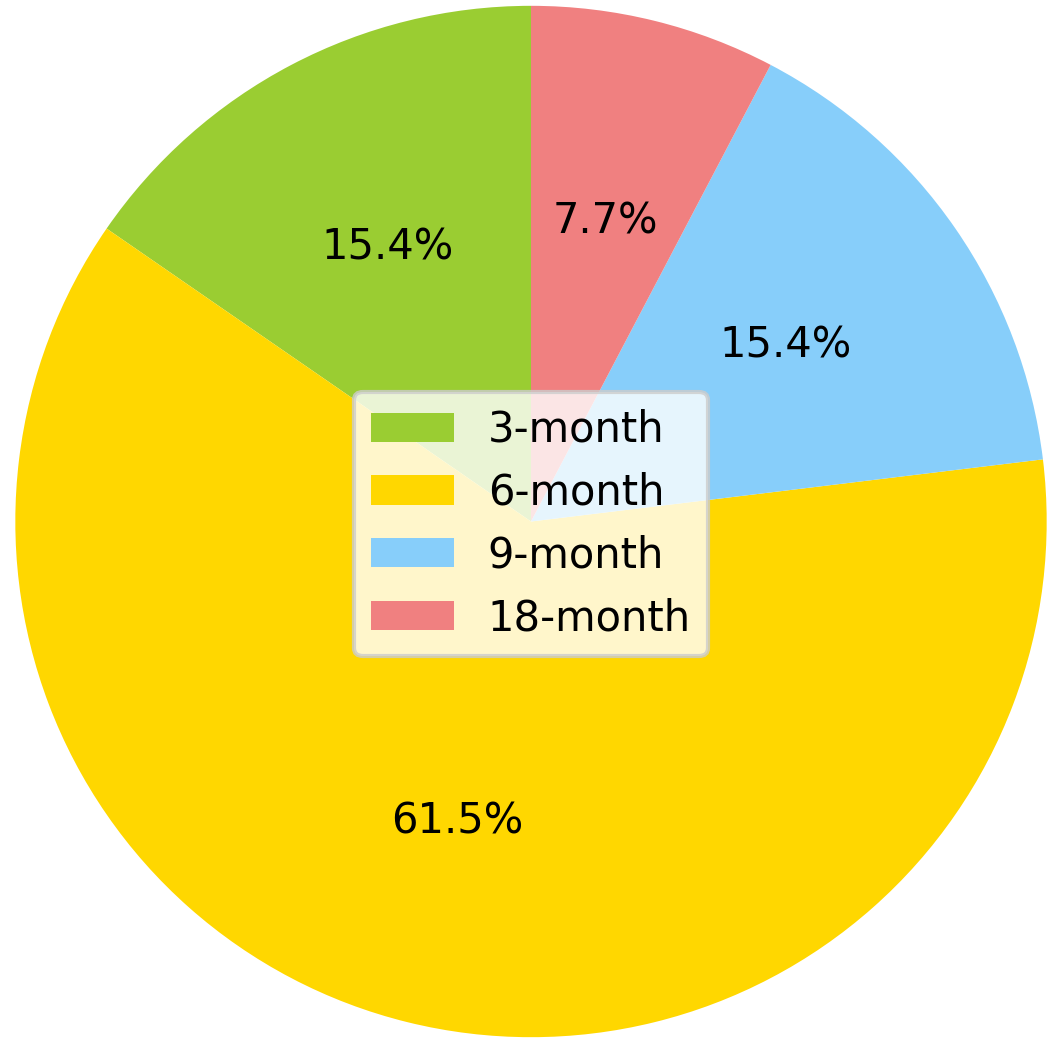}\\
    \end{tabular}
    
    \caption{Left: The distribution data in different classes in InfantSound dataset. Right: the distribution of samples per age of infants.}
    \label{f:dist}
\end{figure}

The dataset consisted of recordings from 15 different families. We did not have transcriptions for all four time periods for each family, so we had less data in the older age ranges. The distribution of data across different vocalizations categories and different ages is shown in Fig~\ref{f:dist}.
\section{Sound Classification Networks}
\label{s:method}
\begin{table}[t]
\caption{Specific architectures of two different CNNs examined in the present study, with or without an Inception Nucleus layer. The convolutional layer parameters are denoted as ''conv (1D or 2D),(number of channels),(kernel size),(stride)".}
    \centering
    \scalebox{0.80}{
\begin{tabular}{|c|c|}
     \hline
     \multicolumn{2}{|c|}{Neural Networks Configurations} \\
     \hline
     CNN with Inception & CNN without Inception \\
     \hline
     302 K & 540 K \\
     \hline
     \multicolumn{2}{|c|}{Input ($8000 \times 1$)} \\
     \hline
     Conv1D,32,80,4&-\\ 
     \hline
     -&Conv1D,32,9,4\\ 
     \hline
     - &Conv1D,32,8,4\\ 
     \hline
     - &Conv1D,32,9,4\\ 
     \hline
     Inception Nucleus:&\\
    Conv1D,64,4,4&\\
     Conv1D,[64,8,4]$\times 2$&-\\
     Conv1D,[64,16,4]$\times 2$&\\
     \hline
     Max Pooling 1D,10,1&Max Pooling 1D,2,1 \\
     \hline
     \multicolumn{2}{|c|}{Reshape (put the channels first)}\\
     \hline
     Conv2D,32,$3\times 3$,1 &Conv2D,64,$3\times 3$,1\\
     \hline
     \multicolumn{2}{|c|}{Max Pooling 2D,$2\times 2$,2}\\
     \hline
     Conv2D,64,$3\times 3$,1&Conv2D,128,$3\times 3$,1\\
     Conv2D,64,$3\times 3$,1&Conv2D,128,$3\times 3$,1\\
     \hline
     \multicolumn{2}{|c|}{Max Pooling 2D,$2\times 2$,2}\\
     \hline
     Conv2D,128,$3\times 3$,1 &Conv2D,256,$3\times 3$,1\\
     \hline
     \multicolumn{2}{|c|}{Max Pooling 2D, $2\times 2$,2}\\
     \hline
     Conv2D,10,$3\times 3$,1 &Conv2D,10,$1\times 1$,1\\
     - &Conv2D,10,$1\times 1$,1\\
     \hline
     \multicolumn{2}{|c|}{Flattening} \\
     \hline
     \multicolumn{2}{|c|}{Softmax} \\
     \hline
\end{tabular}}
\label{t:Net}
\end{table}
\begin{table*}[t]
\caption{Performance of CNNs on different subsets of the infant vocalization dataset. The first column denotes the experiment, the second column reports the results of the CNN with Inception, the third column reports the CNN without Inception Nucleus layer and the last column shows the performance at chance.}
\centering
 \begin{tabular}{c c c c c} 
 \hline
Model Comparison & CNN with Inception&CNN without Inception & Chance\\ [0.5ex] 
 \hline\hline
Infant vs. Adult& $94.12\%$& $97.01\%$& $50.00\%$ \\
\hline
Vocalization vs. Non-Vocalization& $90.49\%$& $82.13\%$&$50.00\%$ \\
\hline
Canonical vs. Non-Canonical& $76.17\%$& $77.03\%$&$50.00\%$ \\
\hline
IDS vs. ADS & $52.72\%$&$60.28\%$& $50.00\%$ \\
\hline
Laugh/Cry vs. Can./Non-Can. vs. IDS/ADS & $98.09\%$& $97.86\%$&$33.33\%$ \\
\hline
Laugh/Cry vs. Can. vs. Non-Can. vs. IDS/ADS & $74.69\%$& $64.27\%$&$25.00\%$ \\
\hline
Laugh/Cry vs.  Can. vs. Non-Can. vs. IDS vs. ADS & $64.32\%$& $41.34\%$&$20.00\%$ \\
\hline
\end{tabular}
\label{t:perf}
\end{table*}
We investigated two end-to-end neural networks that take time-domain waveform data as inputs and process them through several 1D and 2D convolutions to map onto the desired output classes. The main difference between the two networks was the inclusion or exclusion of an inception nucleus. The model architectures are diagrammed in in Table~\ref{t:Net}, and the main parts of both models are summarzied below, including the inception nucleus.

\vspace{0.15cm}
\textbf{Convolution Layers.} Convolutional networks are designed to learn filters that are passed over spatial or temporal dimensions, such as images or acoustic waveforms. Convolutional filters help models learn translation-invariant features and they are efficient in terms of number of parameters that need to be trained~\cite{ebrahimpour_attention}. Convolutional layers can also be stacked to learn representations that build on each other, from small, low-level features to larger, high-level objects~\cite{ebrahimpour_ijcnn}. Finally, convolutional layers can handle inputs of varying lengths and features of varying scales~\cite{ebrahimpour_caption}.

The input layer to our network is a 1D array, representing the audio waveform, which is denoted as $X \in \mathbb{R}^{8000 \times 1}$, since the audio files are 1 second, and the sampling rate was set to be $8\ \mbox{\it kHz}$. The network is designed to learn a set of parameters, $\omega$, to map the input to the prediction, $\hat{Y}$, based on nested mapping functions, given by Eq~\ref{e:net}. 
\begin{equation}
\label{e:net}
\hat{Y} = F(X|\omega) = f_k(...f_2(f1(X|\omega_1)|\omega_2)|...\omega_k)
\end{equation}
where $k$ is the number of hidden layers and $f_i$ is a typical convolution layer followed by a pooling operation. 

\vspace{0.15cm}
\textbf{Inception Nucleus Layer.}
We use an inception nucleus~\cite{ebrahimpour_icassp} for more robust classification and reduced sensitivity to idiosyncratic variance in audio files. The inputs to the inception nucleus are the feature maps of the previous layer, and the nucleus itself consisted of three 1D convolutions with different kernels that are simultaneously applied to inputs in order to capture a range of features and scales. Kernel sizes have significant impacts on the performance of end-to-end architectures for sound classification tasks. The receptive fields of the resulting feature maps are concatenated in a channel-wise manner. By using the Inception Nucleus, the network is given multiple convolutional layers with a variety of kernel sizes to capture richer features without additional tuning of hyperparameters. 

\vspace{0.15cm}
 \textbf{Reshape.}
After applying 1D convolutions on the waveforms to obtain low-level features, the feature map, $L$, will be $\in \mathbb{R}^{1 \times m \times n}$. We can treat $L$ as a grayscale image with width=$m$, height=$n$, and channel=$1$. For simplicity, we transpose the tensor $L$ to $L^{\prime} \in \mathbb{R}^{m \times n \times 1}$. From here, we apply normal 2D convolutions with the VGG~\cite{vgg} standard kernel size of $3 \times 3$ and stride = 1~\cite{vgg}. Also, the pooling layers have kernel sizes = $2 \times 2$ and stride = 2. By converting the acoustic waveform into a learned image representation, we are able to borrow deep learning techniques from the image processing literature, described next.

\vspace{0.15cm}
\textbf{Conv2D.} After the reshape layer, the receptive field can be treated as a grayscale image. We applied several 2D convolution layers similar to those used in the VGG network frequently used in the computer vision literature~\cite{vgg}. The kernel size and stride of the 2D convolutions was fixed to $3 \times 3$ and $1$, respectively. Each 2D convolutional layer was followed by a max pooling layer with a kernel size of $2 \times 2$ and stride of $2$. 

\vspace{0.15cm}
 \textbf{Flatten and Fully Connected layers.} The last convolution layer is followed with a co-called ``flatten'' layer which collapses the spatial dimensions to a single column vector that is passed to a fully connected layer. The number of units in the fully connected layer was set to the number of classes (which varied across models, see below) and the activation function of the final output layer was the \emph{softmax} function.

\begin{table*}[t]
\caption{Classification accuracy is shown for the CNN with Inception Nucleus  on three different comparisons as a function of age. $\#$ indicates the number of testing samples.}
\centering
 \begin{tabular}{c c | c c | c c| c c |c} 
 \hline
 $\#$& 3-month & $\#$ &6-month  & $\#$& 9-month & $\#$&18-month&Model Comparison\\ [0.5ex] 
 \hline\hline
 228&71.05& 219&84.93 &15 & 100.00& 72&95.83 &Non-Vocalization vs. Vocalization \\ 
\hline
 72&100.00&93&96.77&477&91.19&69&68.12&Canonical vs. Non-Canonical \\
\hline
 123&53.16& 210&62.86 & 15&60.00 & 63&63.37&Infant Directed vs. Adult Directed\\
\hline
\end{tabular}
\label{t:perf_based_on_age}
\end{table*}
\section{Model Results}
\label{s:experiments}
Models were trained on a total of 2456 audio clips, each being one second long. Recordings were down-sampled to $8\ \mbox{\it kHz}$ and standardized to zero mean and unit variance. We shuffled the training data to enhance variability in the training set, and trained models using the Adam~\cite{adam} optimizer. The optimizer is a variant of stochastic gradient descent that adaptively tunes the step size for each dimension. We used Glorot weight initialization~\cite{glorot} and trained each model with batch size $128$ for up to $300$ epochs until convergence. To avoid overfitting, all weight parameters were penalized by their $\ell_2$ norm, using a $\lambda$ coefficient of $0.0001$. Our models were implemented in Keras~\cite{keras} and trained using a GeForce RTX 2080 GPU.\\ %

Investigation of model parameters showed that bigger kernel sizes in the first layer lead to features that are more useful for classification, so larger sizes are reported in the following sections. Also, we found that deeper networks with larger numbers of parameters were less able to generalize learning to novel testing sets. Interestingly, this latter finding runs counter to results from the image recognition literature, in which deeper networks tend to generalize better than shallower ones~\cite{resnet,densenet,senet}. The detriment of additional hidden layers may be attributable to the limited number of training examples, which can be tested in future studies with larger datasets. For the present study, we focused on smaller, more shallow networks with relatively fewer parameters.

\vspace{0.15cm}
\textbf{Adult Vocalizations vs. Infant Sounds.}
We combined the classes of canonical, non-canonical, and laugh/cry together to represent the ``Infant'' class, and combined IDS and other adult speech to represent the ``Adult'' class. We then trained our neural networks to distinguish between these two classes. The results are illustrated in the first row of Table~\ref{t:perf}. The model learned to discriminate these two perceptually distinct categories without difficulty and regardless of the use of an inception nucleus.  

\vspace{0.15cm}
\textbf{Infant Vocalizations vs. Non-Vocalizations.}
Next we tested an ostensibly more challenging distinction between two broad classes of infant sounds. The ``Vocalization'' class contained canonical and non-canonical babbling whereas the ``Non-Vocalization'' class contained crying and laughing sounds. We trained both neural networks on the corresponding subset of training examples, and results are shown in the second row of Table~\ref{t:perf}. Performance dropped by 4-5\% compared with the adult vs.\ infant model, which is consistent with the intuition that it more challenging to separate sounds made by infants. Also, the inception nucleus added eight percentage points to performance, indicating its potential utility in perceptually more challenging tasks.

\vspace{0.15cm}
\textbf{Canonical vs. Non-Canonical Babbling.}
Next we tested an even finer-grained distinction between two basic kinds of infant vocalizations. Both classes contain vowels and consonants, with the main difference being their quality relative to adult speech and their sequential structure, or lack thereof. As shown in the third row of Table~\ref{t:perf}, model performance was still well above chance, but it dropped considerably compared with the prior two models, as expected. Interestingly, no benefit was conferred by the inception nucleus for distinguishing between different kinds of babbling. The two classes in this model were more homogeneous compared with classes in the previous two models, suggesting that the greater range of convolutional filters in the inception nucleus is more useful for representing relatively broad and diverse classes.

\vspace{0.15cm}
\textbf{Infant Directed Speech vs.\ Adult Directed Speech.}
The models tested so far included classes that can be distinguished perceptually, based on gross differences in the properties of their sounds. Next we tested a distinction in vocalizations that is much more subtle in terms of differences in acoustic properties. In particular, we tested adult vocalizations that were categorized as either directed or not directed at the infant. So-called ``motherese'', a.k.a.\  infant-directed speech, tends to be more variable in intonation and intensity, and exaggerated in articulation, compared with adult-directed speech. However, these acoustic cues are highly variable within and between speakers, and it can be difficulty even for human listeners to perceive the difference. Model results are shown in the forth row of Table~\ref{t:perf}, and the expected difficulty of this distinction was born out in near-chance performance with the inception nucleus. Surprisingly, the model performed better without the inception nucleus, but recall that the inception nucleus was also not beneficial for distinguishing the two types of babbling. Two types of babbling and two types of adult speech are all relatively homogeneous categories, again suggesting that the greater range of convolutional filters in the inception nucleus is more useful for representing relatively broad and diverse classes.

\vspace{0.15cm}
\textbf{Multiple Category Classification.}
The previous models that tested binary distinctions were useful for investigating the model in terms of its capability to learn different perceptual categories, and the effect of including the inception nucleus. However, for both theory and application, it is important to demonstrate an ability to distinguish multiple classes simultaneously. We ran three models that were trained to learn from three to five different classes simultaneously, as shown in the last three rows of Table~\ref{t:perf}. Results showed that performance was well above chance for all three models, and the inception nucleus was more beneficial as the number of classes increased, which corresponded with an increase in the diversity of sounds that needed to be classified.

\vspace{0.15cm}
\textbf{Generalization on Samples Outside of our Training Set.} Results presented thus far demonstrate the ability of CNN neural networks to learn a variety of classes of infant and adult vocalizations. In the models reported, training and testing sets were drawn from the same 15 families, which raises the question of whether learning would generalize to infants and adults not in the training set. We ran another set of models in which we tested performance by excluding recordings from one family in the training set, and testing performance on the untrained family. Performance dropped only slightly when generalizing to sounds from untrained infants and adults, by 3\% at most. Therefore the model architecture and size and diversity of the data set enabled robust generalization. 

\vspace{0.15cm}
\textbf{Classification Performance as a Function of Age.}
Distinguishing different types of infant and adult vocalizations may be useful for infant monitoring applications, and analyses of learned representations may provide a data-driven method for studying how such complex sounds are processed and produced by human perceptual and motor systems. These analyses are left to future research, but our data set affords a relatively simple, initial test of the model as a tool for studying speech development. In particular, we can test performance as a function of infant age, and patterns of performance over time may provide information about changes in the sound structures of different vocalizations. 

We examined classification performance for the CNN with Inception as a function of age for the three binary distinctions reported above. As shown in Table~\ref{t:perf_based_on_age}, results varied as a function of age, and the effect of age was different depending on the classes being distinguished. The strongest effect of age was for infant vocalizations versus non-vocalizations that became more distinct from each other as infants grew older. This result is consistent with the hypothesis that vocalizations become more clearly speech-like over the course of development. Interestingly, the opposite pattern was found for canonical versus non-canonical babbling. As speech develops, babbling becomes more and more canonical because it becomes more and more speech-like. Thus model results may have tracked the gradual extinction of non-canonical babble over the course of speech development. Finally, adult vocalizations were most difficult to discriminate when the infant was youngest at 3 months of age, suggesting that adults are less engaged in infant-directed speech when infants are less responsive at such an early age. 

\section{Conclusion}
In this study, we developed, optimized, and tested CNNs with and without Inception Nucleus components, up to 17 layers deep, on end-to-end classification of infant sounds and vocalizations that were recorded in natural settings along with vocalizations and vocal interactions of caregivers and adults. Our results contribute to the machine learning community as well as the developmental speech science community by showing how deep learning techniques developed for speech and image recognition can be leveraged for speech development. Further research is needed to control for the number of training and testing examples per class, and confirm that results as a function of age were not unduly influenced by differences in the numbers of training and testing examples. Further work is also needed to test whether the model may shed light on differences between normal and disordered speech development, and whether important aspects of these differences might be revealed through comparisons of learned representations in models trained on speech from different populations. We believe that models like those presented in this paper are opening new avenues toward diagnostic tools for measuring normal vs.\ abnormal speech development, as well as research tools for examining the developmental time course of speech.

\section{Acknowledgements}
We would like to thank Dr.~Anne Warlaumont for gathering the original recordings, and Dr.~Gina Pretzer for collecting and providing the annotations. We would also like to thank our many research assistants who contributed to this project. This research was supported in part by NSF Award 1529127 and a gift from Accenture Labs (PI Kello).

\bibliographystyle{IEEEtran}
\bibliography{mybib}

\end{document}